\documentclass[useAMS,usenatbib]{mn2e}
\usepackage{times,graphicx,float}
\usepackage{psfig}

\title[Universal variabilities in GRBs and blazars]
  {The extension of variability properties in gamma-ray bursts to blazars}
\author[Wu et al.]{Qingwen Wu$^{1}$, Bing Zhang$^{2,3,4}$, Wei-Hua Lei$^{1}$, Yuan-Chuan Zou$^{1}$, En-Wei Liang$^{5,6}$, Xinwu Cao$^{7,8}$  \\
 $^1$School of Physics, Hashing University of Science and Technology, Whang 430074, China; Corresponding author, E-mail: qwwu@hust.edu.cn\\
 $^2$Department of Physics and Astronomy, University of Nevada Las Vegas, NV 89154, USA\\
 $^3$Department of Astronomy, School of Physics, Peking University, Beijing 100871, China\\
 $^4$Kavli Institute for Astronomy and Astrophysics, Peking University, Beijing 100871, China\\
 $^5$Department of Physics and GXU-NAOC Center for Astrophysics and Space Sciences, Guangxi University, Nanning, 530004, China\\
 $^6$Guangxi Key Laboratory for the Relativistic Astrophysics, Nanning 530004, China\\
 $^7$ SHAO-XMU Joint Center for Astrophysics, Shanghai Astronomical Observatory, Chinese Academy of Sciences, 80 Nandan Road, Shanghai 200030, China\\
 $^8$Key Laboratory of Radio Astronomy, Chinese Academy of Sciences, 210008 Nanjing, China\\}

\date{}

\pagerange{\pageref{firstpage}--\pageref{lastpage}} \pubyear{0000}

\def\LaTeX{L\kern-.36em\raise.3ex\hbox{a}\kern-.15em
    T\kern-.1667em\lower.7ex\hbox{E}\kern-.125emX}

\newcommand{\be}{\begin{equation}}
\newcommand{\ee}{\end{equation}}

\newcommand{\msun}{{M}_{\sun}}

\begin{document}

\maketitle

\begin{abstract}
   Both gamma-ray bursts (GRBs) and blazars have relativistic jets pointing at a small angle from our line of sight. Several recent studies suggested that these two kinds of sources may share similar jet physics. In this work, we explore the variability properties for GRBs and blazars as a whole. We find that the correlation between minimum variability timescale (MTS) and Lorentz factor, $\Gamma$, as found only in GRBs by Sonbas et al. can be extended to blazars with a joint correlation of $\rm MTS\propto\Gamma^{-4.7\pm0.3}$. The same applies to the $\rm MTS\propto \it L_{\gamma}^{\rm -1.0\pm0.1}$ correlation as found in GRBs, which can be well extended into blazars as well. These results provide further evidence that the jets in these two kinds of sources are similar despite of the very different mass scale of their central engines. Further investigations of the physical origin of these correlations are needed, which can shed light on the nature of the jet physics.
\end{abstract}


\section{Introduction}
  Relativistic jets are common in black-hole (BH) systems, where the BH mass ranges from stellar mass ($\sim10\msun$, $\msun$ is solar mass, e.g., in BH X-ray binaries, XRBs, and gamma-ray bursts, GRBs) to supermassive scale ($\sim 10^{5-10}\msun$, e.g., in active galactic nuclei, AGNs). Blazars are an extreme type of AGNs in which the broad band emission is dominated by non-thermal radiation that produced in a relativistic jet pointing towards us. Based on the equivalent width (EW) of emission lines, blazars are divided into two sub-types, i.e., flat spectrum radio quasars (FSRQs) and BL Lacertae objects (BL Lacs): BL Lacs have weak or no emission lines (${\rm EW\le5\AA}$), whereas FSRQs have strong emission lines \citep[${\rm EW>5\AA}$, e.g.,][]{up95}, where the different emission-line properties may be triggered by transition of accretion modes \citep[e.g.,][]{gc01,wu08,xu09,wu13}. The spectral energy distribution (SED) of blazars generally include two broad humps in the $\log\nu$-$\log (\nu F_\nu)$ diagram: a low energy hump peaking at infrared to soft X-ray band that is produced by synchrotron radiation, and a high energy hump peaking at MeV to GeV band that is normally attributed to inverse Compton (IC) scattering. GRBs are the most energetic explosions in the universe with a total isotropic-equivalent energy $E_{\rm iso}\sim 10^{49-55}$ erg \citep[e.g.,][for reviews and the references therein]{ma06,zh11}, which may be caused by either collapses of massive stars or mergers of compact objects. The emission of a GRB is believed to originate from an ultra-relativistic jet with a Lorentz factor of a few hundreds. Such a high Lorentz factor is evidenced from the ``compactness" argument \citep[e.g.,][]{fe93,zou11} or the early afterglow light curves that show the signal of fireball deceleration \citep[e.g.,][]{sa99,li10,gh12}. The prompt emission of GRBs in the sub-MeV energy range normally exhibits a non-thermal nature with a smoothly-joint broken power-law shape \citep[i.e. the ``Band function",][]{ba93}. A thermal component superposed on the non-thermal component was also observed in some bursts. The radiation mechanism of GRBs is still hotly debated \citep[e.g., synchrotron, inverse Compton, thermal emission or a mix of them, see][for recent reviews and references therein]{zh14,kz14,pe15}.

  It is believed that the jet physics (e.g., formation mechanism, energy dissipation, radiation process etc.) in BH systems should be intrinsically similar, which is supported by the so-called ``fundamental plane'' as found in XRBs and AGNs \citep[e.g.,][]{mhd03,fa04,do14}. Despite the large discrepancy of the masses of AGNs and GRBs, several universal correlations have been found among the two types of objects, strengthening this possibility. For example, \citet{ww11} found that the GRB \emph{afterglows} and BL Lacs may have the same radiation mechanism based on a similar relation in radio luminosity and spectral slope in radio and optical band, which is further confirmed by \citet{wa14} with two new similar relations of spectral properties. \citet{wu11} found a universal correlation between synchrotron luminosity and Doppler factor in GRBs and blazars, where they assumed the \emph{prompt} emission of GRBs is dominated by synchrotron emission. \citet{ne12} and \citet{zj13} found the efficiency of energy dissipation of jets is similar in GRBs and blazars. A similar case was also made for XRBs and low-luminosity AGNs \citep{ma14}. \citet{lyu14} tried to explore the possible unified radiation physics in GRBs and blazars.

\begin{table*}
\begin{minipage}{170mm}
\footnotesize
  \centerline{\bf Table 1. The GRB sample}
  \begin{tabular}{lccccc||lccccc}
  \hline\hline
Name & $z$ & $\log \rm MTS $ & $\Gamma$ & $\log L_{\gamma}$ & Refs. & Name & $z$ & $\log\rm MTS$ & $\Gamma$ & $\log L_{\gamma}$ & Refs  \\
\hline
060210       & 3.91  &  -0.59$\pm$0.07   & $264_{-4}^{+4}$    &  51.97  & 1,3 & 090510      & 0.90  &  -2.58$\pm0.09$   & 750$_{-50}^{+50}$   &  53.61   & 2,5  \\
060607A      & 3.08  &   0.24$\pm$0.07   & $296_{-8}^{+28}$   &  51.57  & 1,3 & 090618      & 0.54  &  -0.73$\pm0.41$   &   ...               &  52.31   & 2   \\
061007       & 1.26  &  -1.15$\pm$0.08   & $436_{-3}^{+3}$    &  52.50  & 1,3 & 090812      & 2.45  &  -0.58$\pm0.07$   & 501                  &  52.27  & 1,6 \\
061121       & 1.31  &  -1.28$\pm$0.07   & $175_{-2}^{+2}$    &  51.88  & 1,3 & 090902B     & 1.82  &  -2.50$\pm0.15$   & 750$_{-150}^{+150}$  &  53.62  & 2,5  \\
070318       & 0.84  &   0.43$\pm$0.07   & $143_{-7}^{+7}$    &  50.62  & 1,3 & 090926A     & 2.11  &  -2.03$\pm0.06$   & 800$_{-200}^{+200}$  &  53.87  & 2,5  \\
071010B      & 0.95  &  -0.34$\pm$0.07   & $209_{-4}^{+4}$    &  51.15  & 1,3 & 091003      & 0.90  &  -1.67$\pm0.12$   &   ...                &  51.93  & 1   \\
071031       & 2.69  &   0.73$\pm$0.08   & $133_{-3}^{+17}$   &  50.98  & 1,3 & 091029      & 2.75  &  -0.17$\pm0.08$   &  221                 &  51.85  & 1,6  \\
080319B      & 0.94  &  -1.68$\pm$0.11   &   ...              &  52.65  & 1   & 100414      & 1.37  &  -1.96$\pm0.14$   &  ...                 &  52.75  & 2   \\
080319C      & 1.95  &  -0.22$\pm$0.07   & $228_{-5}^{+5}$    &  52.35  & 1,3 & 100621A     & 0.54  &  -0.09$\pm0.07$   &  52                 &  51.04   & 1,6  \\
080804       & 2.21  &  -0.66$\pm$0.23   &   ...              &  52.43  & 2   & 100728B     & 2.11  &  -1.25$\pm0.08$   &  373                 &  51.89  & 2,6  \\
 080810      & 3.35  &  -1.64$\pm0.18$   &409$_{-34}^{+34}$   &  52.08  & 2,3 & 100906A     & 1.73  &  -0.63$\pm0.08$   &  369                 &  51.90  & 1,6  \\
 090102      & 1.55  &  -1.89$\pm0.18$   &  440               &  52.94  & 2,4 & 110205A     & 2.22  &  -0.70$\pm0.08$   &  177                 &  51.85  & 1,6  \\
 090323      & 3.59  &  -1.65$\pm0.23$   &   ...              &  53.08  & 2   & 110213A     & 1.46  &  -0.71$\pm0.04$   &  223                 &  51.52  & 2,6  \\
 090424      & 0.54  &  -1.11$\pm0.07$   & 300$_{-79}^{+79}$  &  51.08  & 1,3 & 130427A     & 0.34  &  -1.35$\pm0.07$   &  325$_{-25}^{+25}$  &  53.43   & 1,5  \\
\hline
\end{tabular}
\end{minipage}

\begin{minipage}{170mm}
References: 1) and 2) represent the references for MTS measurements that selected from \citet{gb14} and \citet{go15} respectively.\\
 The Lorentz factors are selected from 3) \citet{lv12}; 4) \citet{gh12}(use reported peak time of afterglow in this work and equation 1 of \citealt{lv12}); 5) \citet{ha15} and 6) \citet{li15}.
\end{minipage}

\end{table*}

\begin{table*}
\begin{minipage}{170mm}
\footnotesize
  \centerline{\bf Table 2. The blazar sample}
  \begin{tabular}{lcccc|lcccc}
  \hline\hline
Name & $z$ & $\log \rm MTS $ & $\Gamma$ & $\log L_{\gamma}$ & Name & $z$ & $\log\rm MTS$ & $\Gamma$ & $\log L_{\gamma}$  \\
\hline
1Jy 0138-097 &  1.03 &  6.03             &                    & $47.02\pm0.05$   & 3C 273       &  0.16 &  4.7              &                & $46.31\pm0.01$   \\
4C +15.05    &  0.41 &  5.78             &  9.9               & $45.82\pm0.06$   & 3C 279       &  0.54 &  5.48             & 20.9           & $47.59\pm0.01$   \\
PKS 0208-512 &  1.00 &  5.61             &                    & $47.54\pm0.02$   & PKS1437+398  &  0.34 &  6.25             &                & $45.30\pm0.13$  \\
3C 66A       &  0.44 &  5.10             &                    & $47.26\pm0.02$   & PKS 1510-089 &  0.36 &  3.84             & 20.7           & $47.40\pm0.01$  \\
AO 0235+164  &  0.94 &  5.94             & 12.1               & $47.95\pm0.02$   & RGB 1534+372 &  0.14 &  6.42             &                & $44.46\pm0.09$  \\
PKS 0336-019 &  0.85 &  6.02             & 23.0               & $46.96\pm0.04$   & 4C +38.41    &  1.84 &  4.81             & 30.5           & $48.72\pm0.00$  \\
PKS 0420-01  &  0.92 &  5.37             & 11.4               & $47.61\pm0.02$   & NRAO 530     &  0.90 &  4.90             & 39.0           & $47.36\pm0.02$  \\
PKS 0440-00  &  0.84 &  4.95             &                    & $47.57\pm0.02$   & 4C +51.37    &  1.38 &  5.70             &                & $47.82\pm0.01$  \\
PKS 454-234  &  1.00 &  4.83             &                    & $48.13\pm0.01$   & 1Jy 1749701  &  0.77 &  4.68             &  7.5           & $46.82\pm0.03$  \\
4C -02.19    &  2.29 &  5.73             & 16.2               & $48.01\pm0.05$   & OT 081       &  0.32 &  5.47             &                & $46.18\pm0.02$  \\
PKS 0521-36  &  0.06 &  4.57             &                    & $44.68\pm0.02$   & S5 1803+784  &  0.68 &  4.95             &  9.5           & $47.07\pm0.01$  \\
PKS 0537-286 &  3.10 &  6.21             &                    & $48.59\pm0.03$   & B2 1811+31   &  0.12 &  6.26             &                & $44.86\pm0.04$  \\
PKS 0537+441 &  0.89 &  6.04             &                    & $48.19\pm0.04$   & 4C +56.27    &  0.66 &  6.60             & 37.8           & $46.93\pm0.02$  \\
PKS 0716+714 &  0.30 &  4.80             & 10.3               & $46.73\pm0.01$   & PKS 1830-210 &  2.51 &  4.44             &                & $49.35\pm0.00$  \\
PKS 0735+17  &  0.42 &  6.05             &                    & $46.58\pm0.02$   & 1Jy 2005-489 &  0.07 &  6.48             &                & $44.77\pm0.03$  \\
PKS 0829+046 &  0.17 &  6.06             &                    & $45.61\pm0.02$   & PKS 2052-47  &  1.49 &  5.57             &                & $48.28\pm0.01$  \\
4C +71.07    &  2.17 &  4.41             & 28.0               & $48.35\pm0.03$   & 4C -02.81    &  1.29 &  5.88             &                & $47.17\pm0.05$  \\
OJ 287       &  0.31 &  5.11             & 15.4               & $46.12\pm0.02$   & S3 2141+17   &  0.21 &  5.45             &                & $46.02\pm0.01$  \\
S4  0917+44  &  2.19 &  5.08             &                    & $48.67\pm0.01$   & PKS 2155-304 &  0.12 &  6.69             &                & $45.99\pm0.01$  \\
4C +55.17    &  0.90 &  5.50             &                    & $47.60\pm0.02$   & BL Lac       &  0.07 &  4.96             & 11.6           & $45.10\pm0.01$  \\
S4 0954+658  &  0.37 &  5.67             &                    & $45.98\pm0.03$   & B3 2247+381  &  0.12 &  6.15             &                & $44.68\pm0.07$  \\
4C +29.45    &  0.72 &  5.59             & 25.1               & $47.30\pm0.01$   & 3C 454.3     &  0.86 &  2.94             & 19.9           & $48.74\pm0.01$  \\
4C +21.35    &  0.43 &  3.64             & 45.5               & $47.44\pm0.01$                                                                                 \\
\hline
\end{tabular}
\end{minipage}

\end{table*}

  Both GRBs and blazars are highly variable sources. The typical minimum variability timescale (MTS) for blazars is around one day \citep[e.g.,][]{vo13}, where ultra-fast variability as short as 3-5 min at TeV energies is also found in some blazars \citep[e.g.,][]{ah07,al07,al14}. The typical MTS of GRBs is around 0.5 s with the shortest timescale on the order of 10 ms \citep[e.g.,][]{ma13,gb14}, where the MTS is much shorter than the overall duration (e.g., $T_{90}$). The physical origin of the flux variation is unclear for both GRBs and blazars, where several models have been proposed to explain their temporal variability. The conventional models such as the internal shock scenario \citep[e.g.,][]{ko97} and the photospheric scenario \citep[e.g.,][]{la09} link the rapid variability of GRBs directly to the activity of the central engine. An alternative scenario envisages that the variability is triggered in the emitting region through local relativistic motion (e.g., local relativistic turbulence or reconnection events), which is not related to the central engine directly \citep[e.g.][]{nk09}. \citet{zy11} suggested that the temporal variability of GRBs may include two variability components: a broad (slow) component related to the central engine activity and a narrow (fast) component associated with relativistic magnetic reconnection and/or turbulence (see e.g. \citealt{zz14}). This scenario can also explain the ultra-fast variability (e.g., several minutes) as found in some blazars at TeV energies \citep[e.g.,][]{ah07,al07,al14}.

  \citet{so14} found that the MTS becomes shorter as the gamma-ray luminosity (or Lorentz factor) increases in GRBs. In this work, we try to extend the gamma-ray variability properties in GRBs to blazars, which is aimed to explore the possible universal variability properties in them and shed light on the physical mechanism that triggered the flux variations in relativistic jets. Throughout this work, we assume the following cosmology: $H_{0}=70\ \rm km\ s^{-1} Mpc^{-1}$, $\Omega_{0}=0.3$ and $\Omega_{\Lambda}=0.7$.

 \section{Sample}
   We select the GRB sample from \citet{so14}, where the correlations between the MTS, Lorentz factors and Gamma-ray luminosities were explored. The MTS of GRBs in \citet{so14} was derived using a Harr wavelet technique, but the MTS was simply defined in reference to a noise floor in the measurement \citep[see also][]{ma13}. \citet{gb14} and \citet{go15} analyzed the MTS from the structure-function method based on non-decimated Haar wavelets for a large sample of GRBs from $Swift$ and $Fermi$. They reported MTS values that are less dependent on the measurement noise level compared former works \citep[e.g.,][]{wa00,ma13,so14}. Since this method get rid of artificial effect due top the noise floor, hereafter we adopt the MTS measurements reported in \citet{gb14} and \citet{go15}. The Lorentz factors reported in \citet{so14} were taken from \cite{lv12} and \citet{gh12}. \cite{gh12} made use of the \citet{bm76} self-similar deceleration solution, which led to typically a factor of 2 times smaller $\Gamma$ than those derived from the conventional method for a same source. Since at the peak time, the blast wave is not in the Blandford-McKee self-similar regime, we believe that the conventional method is more reasonable to estimate $\Gamma$ \citep[see][for more discussions]{lv12}. We therefore select the Lorentz factors that are estimated from the conventional method \citep[e.g.,][]{lv12,li15}. We neglect the sources with only a upper or lower limits for the Lorentz factor, so that the sample is mostly composed of those GRBs whose $\Gamma$ was measured using the conventional afterglow onset method \citep[e.g.][]{sa99}. Four GRBs with Lorentz factors estimated from the GeV data or optical flashes are also selected \citep[][]{ha15}, where the GeV and optical flashes are assumed to be emitted from the external shock. The isotropic mean gamma-ray luminosities of GRBs are selected from \citet{lv12}, which are derived from the total isotropic energy, $E_{\rm iso}$, and burst duration, $T_{90}$, with $L_{\gamma}=(1+z)E_{\rm iso}/T_{\rm 90}$. The redshift, MTS, Lorentz factor, and $L_{\gamma}$ for these GRBs are listed in Table 1. In total, we include 28 GRBs, with 22 having estimated Lorentz factor values.

   A sample of 45 blazars are selected from \citet{vo13}, where the MTS is derived using a structure-function approach for the gamma-ray data from the \emph{Fermi} telescope. The Doppler factor of blazars can be estimated from the timescale and amplitude of radio flares by assuming the variability timescale of the flares correspond to the light-travel time across the emission region and that the intrinsic brightness temperature of the source is limited to the equipartition value \citep[e.g.,][]{ho09}. Combined with the apparent jet speed, it is possible to calculate the Lorentz factor and jet viewing angle for a given blazar. There are 20 blazars with estimated Lorentz factors using this method, which is selected from \citet[for more details and the references therein]{sa10}. The total 0.1-100 GeV $k$-corrected gamma-ray luminosity, $L_{\gamma}$, is calculated from the energy flux ($F_{\gamma}$) and the fitted parameters as reported in \citet{no12}.  The redshift, MTS, Lorentz factor and $L_{\gamma}$ for blazars are reported in Table 2.

 \section{Results}

  \cite{so14} found an anti-correlation between MTS and the Lorentz factor in GRBs, i.e. $\rm MTS\propto \Gamma_0^{-4.1\pm0.6}$ for $\Gamma\ga225$, and a shallow-plateau at smaller Lorentz factors. With the MTS values derived by \citet{gb14} and \citet{go15}, we find a similar $\rm MTS-\Gamma$ anti-correlation for the 22 GRBs with estimated Lorentz factors, with no evidence of the shallow-plateau (GRB 100621A is an outlier, see Figure 1). After excluding GRB 100621A, a best straight-line fit to GRB data with errors\footnote{We adopt a linear fitting that considers the uncertainties in both $x$- and $y$- coordinates provided in NUMERICAL RECIPES IN FORTRAN in all the linear fittings in this work (fitxy.for, http://www.nrbook.com/a/bookfpdf.php).} in both $\Gamma$ and MTS gives (solid line in Figure 1)
   \be
  \log \frac{\rm MTS}{1+z}=-(4.8\pm1.5) \log \Gamma + (11.1\pm3.2),
  \ee
  where a 50\% error is adopted for $\Gamma$ considering uncertainties of the model parameters in deriving them \citep[e.g., one order of magnitude uncertainty in the density of ambient medium and radiative efficiency of the blast wave,][]{wb15,zb15} and 0.2 dex is adopted for MTS considering that the MTS measurements are derived from different telescopes with different wavebands \citep[e.g.,MTS derived from Swift at 15-350 keV maybe 1-1.5 times longer than that derived from Fermi at 8-1000 keV, see][for more details]{go15}.  We find that this anti-correlation as found in GRBs can be extended to blazars, where 20 blazars roughly stay on the best-fit line of GRBs (solid line in Figure 1). Therefore, we further present the best straight-line fit for these 41 sources (20 blazars + 21 GRBs),
   \be
  \log \frac{\rm MTS}{1+z}=-(4.7\pm0.3) \log \Gamma + (10.7\pm0.7),
  \ee
  where a typical uncertainty of 0.4 dex and 0.2 dex is assigned for MTS of blazars \citep[e.g.,][]{vo13} and GRBs \citep[e.g.,][]{go15}, respectively, and a 50\% or 30\% error is introduced for $\Gamma$ for GRBs \citep[e.g.,][]{lv12,so14} and blazars \citep[e.g.,][]{ho09}, respectively. It is found that the global slope for blazars/GRBs is roughly consistent with that found in 21 GRBs only within uncertainties.

     \begin{figure}\label{fig1}
   \centering
   \includegraphics[width=8cm]{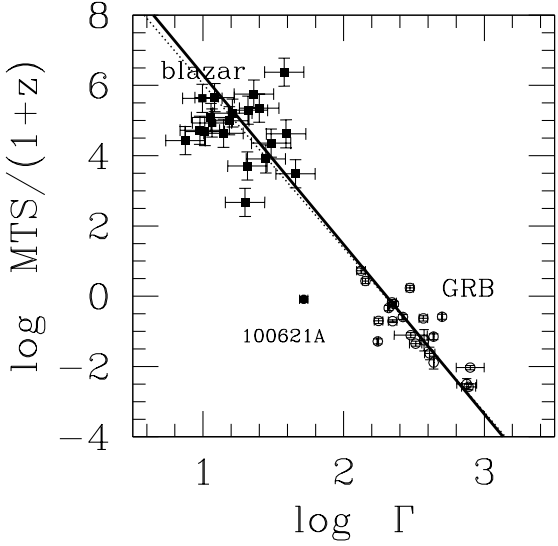}
   \caption{The minimum variability timescale, MTS, vs Lorentz factor, $\Gamma$, for GRBs and blazars. The solid line represents the best linear fit for 21 GRB (open circles, excluding a GRB 100621A that deviate the correlation evidently), which can roughly extended to blazars. The dotted line represents the best fit for GRBs (excluding GRB 100621A) and blazars in our sample. }
  \end{figure}

   \begin{figure}\label{fig2}
   \centering
   \includegraphics[width=8cm]{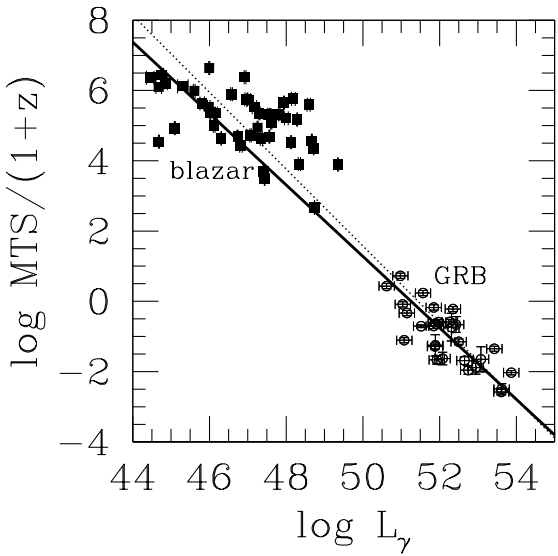}
   \caption{MTS vs gamma-ray luminosity, $L_{\gamma}$, for GRBs and blazars. The solid line represents the best linear fit for GRB sources, where blazars roughly stay on the best-fit line of GRBs. The dotted line represents the best fit for blazars (squares) and GRBs(circles). }
  \end{figure}

   \cite{so14} found that MTS is also anti-correlated with gamma-ray luminosity, $L_{\gamma}$, in GRBs, where the relation also becomes flat when $L_{\gamma}\la10^{51}\rm erg\ s^{-1}$. With the MTS values adopted from \citet{gb14} and \citet{go15}, we find a similar anti-correlation, where the best straight-line fit to GRB data with errors in both $L_{\gamma}$ and MTS gives (solid line in Figure 2)
    \be
  \log \frac{\rm MTS}{1+z}=-(1.0\pm0.1) \log L_{\gamma} + (51.2\pm3.8),
  \ee
   where a typical uncertainty of 0.2 dex is adopted for both MTS \citep[e.g.,][]{go15} and $L_{\gamma}$ \citep[e.g.,][]{lv12} for these GRBs.
   The blazars also roughly stay on the extension of the best-fit line of GRBs (solid line in Figure 2).  We also give the best straight-line fit to data of GRBs and blazars considering the uncertainties of both quantities, which gives
   \be
  \log \frac{\rm MTS}{1+z}=-(1.1\pm0.1) \log L_{\gamma} + (56.4\pm0.7),
  \ee
   where the typical uncertainty of 0.4 dex is adopted for the MTS of blazars \citep[e.g.,][]{vo13}. Again the slope of GRBs/blazars is quite consistent with that derived in pure GRBs.

  \section{Conclusion and Discussion}

   GRBs and blazars are two populations of sources with the most powerful and relativistic jets in the universe. Compared with the accretion physics, the jet physics is less clear. The MTS provides information on the size and location of gamma-ray emitting region, which may shed light on the physical process responsible for the variability (e.g., central engine activity, the Lorentz factor of the jet, and probably local relativistic reconnection/turbulence in the jet). In this work, we explore the possible universal variability properties in GRBs and blazars by compiling 28 GRBs and 45 blazars with measured MTS, gamma-ray luminosities and estimated Lorentz factors from the literature. We find the correlations of $\rm MTS-\Gamma$ and $\rm MTS-L_{\gamma}$ as found in GRBs can be naturally extended to blazars, which not only provide further evidence on the similarities of the jets in GRBs and blazars but also give possible new clues to understand jet physics in these BH systems.

   The slope of $\rm MTS-\Gamma$ correlation is $-4.7\pm0.3$ for GRBs and blazars, which is quite consistent with $-4.8\pm1.5$ as found in pure GRB sample. Incidentally, the slope of $\rm MTS-L_{\gamma}$ for the entire GRB-blazar sample ($-1.1\pm 0.1$) is also quite consistent with that found in GRBs only ($-1.0\pm0.1$). The similar slopes of the GRB/blazar sample and the pure GRB sample strengthen that the correlations as found in GRBs can be extended to blazars. These results provide new similarities on the jet physics in GRBs and blazars, where several other similar properties have been found in last several years\citep[e.g.,][]{ww11,wu11,ne12}.  It should be noted that \citet{so14} found a knee-like feature in both $\rm MTS-\Gamma$ and $\rm MTS-\it L_{\gamma}$ correlations, where the anti-correlations become flat for $\Gamma\la225$ or  $L_{\gamma}\la 10^{51}\rm erg/s$. This break is not evident in our correlations, with only GRB 100621A having a quite small Lorentz factor ($\Gamma=52$) and evidently deviating from the $\rm MTS-\Gamma$ correlation (it however follows the MTS$-L_{\gamma}$ correlation very well). This is because we adopted the MTS values provided by \cite{gb14} and \cite{go15}, who effectively removed the artificial MTS values near the threshold due to the nose floor. Their wavelet method also ensures a better homogeneity between the GRB and blazar samples, which strengthens our joint analysis.

   There is no evident MTS-$\Gamma$ correlation in the subsample of blazars (correlation coefficient $r=-0.15$ with a probability of $p=$0.5). The MTS and $L_{\gamma}$ is also only mildly correlated ($r=-0.51$ and $p=5\times10^{-5}$) in pure blazars. These correlations are not as tight as those found in pure GRBs ($|r|>0.8$ with $p<10^{-4}$), which prevent us from comparing the slopes of the blazar subsample with those of GRB subsample directly. The poor correlations among blazars may be intrinsic, but could be also caused by some uncertainties or biases in measuring $\Gamma$ and MTS. For example, the $\Gamma$ is derived by combining the measurements of the Doppler factor and apparent jet speed which make use of observational data in different energy bands. The emission from different bands may originate from different emission regions, which may have different Lorentz factor, so that the estimated $\Gamma$ may not be reliable. Also the typical MTS measured for blazars is of the order of several months, the inadequate temporal coverage (e.g. with $Fermi$) may cause observational selection effects. Future data with better quality and more samples are needed to address whether blazar subsample indeed has (or does not have) significant correlations. In any case, the fact that the correlations derived from GRBs can be extended to blazars is intriguing, which may hint towards some common physics behind two types of jets.

   The variability of GRBs and blazars has been historically used to understand the well known ``compactness'' problem, and the minimum bulk Lorentz factors can be estimated if the MTS is used as one of the inputs \citep[e.g.,][]{do95,pi99,li01}. \citet{so14} proposed that the anti-correlation of MTS-$\Gamma$ as found in GRBs can be understood by the opacity effect, even if the slope is different from the $\gamma\gamma$ opacity argument (the gamma-ray luminosity is not included in their analysis). The transparency condition for pair production is $\tau_{\gamma\gamma}<1$, which implies $\rm MTS \propto \it L_{\gamma}/\rm \Gamma^{2\beta+2}$, where $\beta$ is photon index of the spectrum. The high-energy index $\beta$ ranges from  2 to 3 with an average value of 2.6 \citep[e.g.,][]{ta15} in the Band function of GRBs \citep[e.g.,][]{ba93} and $\beta\sim2-3$ for blazars as derived from the indices of electron distribution above the broken electron Lorentz factor in modeling their multi-waveband SEDs \citep[e.g., $p_2\sim3-5$,][]{ka14}. \citet{wu11} found a universal correlation between synchrotron luminosity,  $L_{\rm syn}$, and Doppler factor, $\mathcal{D}$, in GRBs and blazars ($L_{\rm syn}\propto \mathcal{D}^{3.1\pm0.1}$), where the prompt emission of GRBs and the first hump of blazars are assumed to be dominated by synchrotron emission. Assuming gamma-ray luminosities in the second hump is normally more or less similar to $L_{\rm syn}$ of the first hump in blazars and $\mathcal{D}\sim2\Gamma$ for the relativistic jet with viewing angle $\theta\rightarrow 0^{\rm o}$, we find $L_{\gamma}\propto \Gamma^{3.1\pm0.1}$. If this is the case, we find  $\rm MTS \propto \Gamma^{-4.2}$ for the typical value of $\beta\sim2.6$ in blazars and GRBs, which is roughly consistent with that found in GRBs and/or blazars (Eq. 2). It should be cautious that the opacity effect just give the lower limit for the Lorentz factor, which is energy dependent (e.g., depend on the photons at highest energy) and this gives the uncertainties in the above explanation. It is interesting to note that the derived Lorentz factors from the transparency condition in GRBs and blazars are normally just several times lower than that derived from other independent methods \citep[e.g.,][]{zou10,zj12}, which suggests that above explanation based on transparency condition may be reasonable.


  There might be another interpretation to the relation. Based on the observational data, one has the Lorentz factor ratio $\Gamma_{\rm GRB}/\Gamma_{\rm blazar}\sim500/10$ \citep[e.g.,][]{ho09,lv12} and black hole mass ratio $M_{\rm GRB}/M_{\rm blazar}\sim10\msun/10^9\msun$ \citep[e.g.,][]{wa04} between GRBs and blazars. Even though the physics behind this scaling relation is not fully explored, it suggests $M \propto \Gamma^{-4.7}$, i.e. a central engine with a smaller mass launches a faster jet. Since the observed MTS is ${\rm MTS} \propto R/ (\Gamma^2 c) \sim (\Gamma^2 c \delta t_{\rm eng})/(\Gamma^2 c) \sim \delta t_{\rm eng}$, where $t_{\rm eng} \propto 2GM/c^2$ is the minimum variability time scale of the central engine \citep[e.g.][]{ko97,zm04}, one has ${\rm MTS}\propto M$, and hence, $\rm MTS \propto \Gamma^{-4.7}$ (Eq.(2)).
 
 \section*{Acknowledgments}
We thank the anonymous referee for constructive comments and suggestions. This work is supported by the NSFC (grants 11103003, 11133005, 11573009 and U1431124), National Basic Research Program of China (973 Program, No. 2014CB8455800), and New Century Excellent Talents in University (NCET-13-0238).

\label{lastpage}


\begin{thebibliography}{99}


\bibitem[Albert et al.(2007)]{al07} Albert, J., et al. 2007, ApJ, 669, 862

\bibitem[Aleksic et al.(2014)]{al14} Aleksic, J., et al. 2014, Science, 346, 1080

\bibitem[Aharonian et al.(2007)]{ah07} Aharonian, F., et al. 2007, ApJ, 664, 71

\bibitem[Band et al.(1993)]{ba93} Band, D., et al. 1993, ApJ, 413, 281
\bibitem[Blandford \& McKee(1976)]{bm76} Blandford, R. D. \& McKee, C. F. 1976, Phys.Fluids, 19, 1130

\bibitem[Cao(2003)]{ca03}  Cao, X. 2003, ApJ, 599, 147
\bibitem[Dong et al.(2014)]{do14} Dong, A.-J., Wu, Q. \& Cao, X.-F. 2014, ApJ, 787, 20

\bibitem[Dondi \& Ghisellini(1995)]{do95} Dondi, L., \& Ghisellini, G. 1995, MNRAS, 273, 583

\bibitem[Falcke et al.(2004)]{fa04} Falcke, H., K$\ddot{\rm o}$rding, E. G., \& Markoff, S. 2004, A\&A, 414, 895

\bibitem[Fenimore et al.(1993)]{fe93} Fenimore, E. E., Epstein, R. I., \& Ho, C. 1993, A\&AS, 97, 59

\bibitem[Ghirlanda et al.(2012)]{gh12} Ghirlanda, G., Nava, L., Ghisellini, G., Celotti, A., Burlon, D., Covino, S., Melandri, A. 2012, MNRAS, 420, 483

\bibitem[Ghisellini \& Celotti(2001)]{gc01} Ghisellini, G., \& Celotti, A. 2001, ApJ, 379, 1

\bibitem[Golkhou \& Butler(2014)]{gb14} Golkhou, V. Z., \& Butler, N. R. 2014, ApJ, 787, 90

\bibitem[Golkhou et al.(2015)]{go15} Golkhou, V. Z., Butler, N. R., Littlejohns, O. M. 2014, to appear in ApJ (arXiv:1501.05948)

\bibitem[Hascoet et al.(2015)]{ha15} Hascoet, R., Vurm, I., Beloborodov, A. M. 2015, submitted to ApJ (arXiv:1504.06369)

\bibitem[Hovatta et al.(2009)]{ho09} Hovatta, T., Valtaoja, E., Tornikoski, M., \& $\rm L\ddot{a}hteenm\ddot{a}ki$, A. 2009, A\&A, 494, 297

\bibitem[Kang et al.(2014)]{ka14} Kang, S.-J., Chen, L. \& Wu, Q. 2014, ApJS, 215, 5

\bibitem[Kobayashi et al.(1997)]{ko97} Kobayashi, S., Piran, T., \& Sari, R. 1997, ApJ, 490, 92

\bibitem[Kumar \& Zhang(2015)]{kz14} Kumar, P. \& Zhang, B. 2015, Physics Reports, 561, 1

\bibitem[Lazzati et al.(2009)]{la09} Lazzati, D., Morsony, B. J., \& Begelman, M. C. 2009, ApJ, 700, L47

\bibitem[Lei et al.(2013)]{lei13}  Lei, W. H., Zhang, B., \& Liang, E. W. 2013, ApJ, 765, 125

\bibitem[Liang et al.(2010)]{li10} Liang, E.-W., Yi, S.-X., Zhang, J., $\rm L\ddot{u}$, H.-J., Zhang, B.-B., Zhang, B. 2010, ApJ, 725, 2209
\bibitem[Liang et al.(2015)]{li15}  Liang, E.-W., Lin, T.-T., LV, J., Lu, R., Zhang, J., Zhang, B. 2015 (arXiv:1505.03660)

\bibitem[Lithwick \& Sari(2001)]{li01} Lithwick, Y., Sari, R. 2001, ApJ, 555, 540

\bibitem[Lyu et al.(2014)]{lyu14} Lyu, F., Liang, E.-W., Liang, Y.-F., Wu, X.-F., Zhang, J., Sun, X.-N., Lu, R.-J., Zhang, B. 2014, ApJ, 793, 36

\bibitem[L$\rm \ddot{u}$ et al.(2012)]{lv12} L$\rm \ddot{u}$, J., Zou, Y.-C., Lei, W.-H., Zhang, B., Wu, Q. Wang, D.-X., Liang, E.-W., L$\rm \ddot{u}$, H.-J. 2012, ApJ, 751, 49

\bibitem[$\rm M \acute{e} sz \acute{a} ros$(2006)]{ma06} $\rm M \acute{e} sz \acute{a} ros$, P. 2006, Reports on Progress in Physics, 69, 2259

\bibitem[Ma et al.(2014)]{ma14} Ma, R., Xie, F.-G., \& Hou, S. 2014, ApJL, 780, 14

\bibitem[MacLachlan et al.(2013)]{ma13} MacLachlan, G. A., et al. 2013, MNRAS, 432, 857

\bibitem[Merloni et al.(2003)]{mhd03} Merloni, A., Heinz, S., \& Di Matteo, T. 2003, MNRAS, 345, 1057

\bibitem[Narayan \& Kumar(2009)]{nk09} Narayan, R., \& Kumar, P. 2009, MNRAS, 394, L117

\bibitem[Nemmen et al.(2012)]{ne12} Nemmen, R. S., Georganopoulos, M., Guiriec, S., Meyer, E. T., Gehrels, N., \& Sambruna, R. M. 2012, Science, 338, 1445

\bibitem[Nolan et al.(2012)]{no12} Nolan, P. L., et al. 2012, ApJS, 199, 31

\bibitem[Peer(2015)]{pe15} Peer, A. 2015, arXiv:1504.02626

\bibitem[Piran(1999)]{pi99} Piran, T. 1999, Phta. Rep., 314, 575

\bibitem[Sari \& Piran(1999)]{sa99} Sari, R., \& Piran, T. 1999, ApJ, 520, 641

\bibitem[Savolainen et al.(2010)]{sa10} Savolainen, T., Homan, D. C., Hovatta, T., Kadler, M., Kovalev, Y. Y., Lister, M. L., Ros, E., \& Zensus, J. A. 2010, A\&A, 512, 24

\bibitem[Sonbas et al.(2014)]{so14} Sonbas, E., MacLachlan, G. A., Dhuga, K. S., Veres, P., Shenoy, A., Ukwatta, T. N. 2015, ApJ, 805, 86

\bibitem[Tang et al.(2015)]{ta15} Tang, Q.-W., Peng, F.-K., Wang, X.-Y., Tam, P.-H. T. 2015, ApJ, 806, 194

\bibitem[Urry \& Padovani(1995)]{up95} Urry, C. M, Padovani, P. 1995, PASP, 107, 803

\bibitem[Vovk \& Neronov(2013)]{vo13} Vovk, I., Neronov, A.. 2013, ApJ, 767, 103

\bibitem[Wang \& Wei(2011)]{ww11} Wang, J., \& Wei, J. Y. 2011, ApJ, 726, 4

\bibitem[Wang et al.(2014)]{wa14} Wang, F. Y.; Yi, S. X.; Dai, Z. G. 2014, ApJ, 786, 8

\bibitem[Wang et al.(2004)]{wa04} Wang, J.-M.; Luo, B., Luis, C. H. 2004, ApJ, 615, 9

\bibitem[Wang et al.(2015)]{wb15} Wang, X.-G. et al. 2015, ApJS, 219, 9

\bibitem[Walker et al.(2000)]{wa00} Walker, K. C., Schaefer, B. E., Fenimore, E. E. 2000, ApJ, 537, 264

\bibitem[Wu et al.(2013)]{wu13} Wu, Q., Cao, X., Ho, L. C., Wang, D.-X. 2013, ApJ, 770, 31

\bibitem[Wu et al.(2011)]{wu11} Wu, Q., Zou, Y.-C., Cao, X., Wang, D.-X. 2011, Chen, L. ApJ, 740, 21

\bibitem[Wu \& Cao(2008)]{wu08} Wu, Q. \& Cao, X. 2008, ApJ, 687, 156

\bibitem[Xu et al.(2009)]{xu09} Xu, Y.-D., Cao, X., \& Wu, Q. 2009, ApJ, 694, 107

\bibitem[Zhang \& Meszaros(2004)]{zm04} Zhang, B., \& Meszaros, P. 2004, International Journal of Modern Physics A, 15, 2385

\bibitem[Zhang(2011)]{zh11} Zhang, B. 2011, Comptes Rendus Physique, 12, 206

\bibitem[Zhang(2014)]{zh14} Zhang, B. 2014, IJMPD, 23, 1430002

\bibitem[Zhang \& Yan(2011)]{zy11} Zhang, B., \& Yan, H. 2011, ApJ, 726, 90

\bibitem[Zhang \& Zhang(2014)]{zz14} Zhang, B., \& Zhang, B. 2014, ApJ, 782, 92

\bibitem[Zhang et al.(2012)]{zj12} Zhang, J., Liang, E.-W., Zhang, S.-N., Bai, J. M. 2012, ApJ, 752, 157

\bibitem[Zhang et al.(2013)]{zj13} Zhang, J., Liang, E.-W., Sun, X.-N., Zhang, B., Lu, Y., Zhang, S.-N. 2013, ApJ, 774, 5

\bibitem[Zhang et al.(2015)]{zb15} Zhang, B.-B., van Eerten, H., Burrows, D. N., Ryan, G. S., Evans, P. A., Racusin, J. L., Troja, E., MacFadyen, A. 2015, ApJ, 806, 15

\bibitem[Zou et al.(2011)]{zou11} Zou, Y.-C., Fan, Y.-Z., Piran, T. 2011, ApJ, 726, 2

\bibitem[Zou \& Piran(2010)]{zou10} Zou, Y.-C., \& Piran, T. 2010, MNRAS, 402, 1854


\end{thebibliography}
\end{document}